%% file: main.tex
\documentclass[journal]{IEEEtran}
\ifCLASSINFOpdf
\else
\fi
\input{preambol}
\input{imgs_sim}

\input{imgs_real}

\begin{document}
%
\title{Multi-Objective CNN Based Algorithm for SAR Despeckling}
%
%
%

\author{Sergio~Vitale,~\IEEEmembership{Student Member,~IEEE,}
        Giampaolo~Ferraioli,~\IEEEmembership{Senior Member,~IEEE,}
        and~Vito~Pascazio,~\IEEEmembership{Senior~Member,~IEEE}
\thanks{S. Vitale and V. Pascazio  are with Dipartimento di Ingegneria - University of Napoli Parthenope. G. Ferraioli is with Dipartimento di Scienze e Tecnologie - University of Napoli Parthenope.,  e-mails: \{sergio.vitale, giampaolo.ferraioli, vito.pascazio\}@uniparthenope.it }

}

\maketitle

\begin{abstract}
Deep learning (DL) in remote sensing has nowadays become an effective operative tool: it is largely used in applications such as change detection, image restoration, segmentation, detection and classification.
With reference to synthetic aperture radar (SAR) domain  the application of DL techniques is not straightforward due to non trivial interpretation of SAR images, specially caused by the presence of speckle.
Several deep learning solutions for SAR despeckling have been proposed in the last few years. Most of these solutions focus on the definition of different network architectures with similar cost functions not involving SAR image properties.
In this paper, a convolutional neural network (CNN) with a multi-objective cost function taking care of spatial and statistical properties of the SAR image is proposed. This is achieved by the definition of a peculiar loss function obtained by the weighted combination of three different terms.
Each of this term is dedicated mainly to one of the following SAR image characteristics: spatial details, speckle statistical properties and strong scatterers identification.
Their combination allows to balance these effects.
Moreover, a specifically designed architecture is proposed for effectively extract distinctive features within the considered framework.
Experiments on simulated and real SAR images show the 
accuracy of the proposed method compared to the State-of-Art despeckling algorithms, both from quantitative and qualitative point of view.
The importance of considering such SAR properties in the cost function is crucial for a correct noise rejection and details preservation  in different underlined scenarios, such as homogeneous, heterogeneous and extremely heterogeneous.
\end{abstract}


\input{1.introduction}
\input{2.Method}
\input{3.discussion}

\input{4.conclusion}

\appendices
\input{5.appendix}

\IEEEpeerreviewmaketitle
\bibliographystyle{IEEE}
\bibliography{refs}

%

%
%
%




\end{document}

%% file: preambol.tex
\usepackage{pifont}
\usepackage{amsmath,graphicx}
\usepackage{color}
\usepackage{tikz}
\usepackage{balance}
\usepackage{url}
\usepackage{multirow,hhline}
\usepackage{placeins}
\usepackage{wasysym}
\usepackage{amssymb}

\usepackage[nomessages]{fp}
\usepackage{array}

\newcommand{\first}[1]{{\textbf{#1}}}
\newcommand{\second}[1]{{\underline{#1}}}
\newcommand{\cmark}{\ding{51}}%
\newcommand{\xmark}{\ding{53}}%

\newcommand{\hx}{\widehat{X}}
\newcommand{\hn}{\widehat{N}}
\newcommand{\lmse}{\mathcal{L}_2}
\newcommand{\lkl}{\mathcal{L}_{KL}}
\newcommand{\ledge}{\mathcal{L}_\nabla}
\newcommand{\lcost}{\mathcal{L}}

\newcommand{\labsty}[1]{{\footnotesize #1}}
\newcommand{\labRef}{\labsty{Reference}}
\newcommand{\labNoise}{\labsty{Noisy}}
\newcommand{\labProp}{\labsty{MONet}}

\newcommand{\labFans}{\labsty{FANS}}
\newcommand{\labSarbm}{\labsty{SAR-BM3D}}
\newcommand{\labNoland}{\labsty{NOLAND}}
\newcommand{\labIdcnn}{\labsty{ID-CNN}}
\newcommand{\labSardrn}{\labsty{SAR-DRN}}
\newcommand{\labSarcnn}{\labsty{SAR-CNN}}

\newcommand{\labSSIM}{\labsty{\textbf{SSIM}}}
\newcommand{\labMSE}{\labsty{\textbf{MSE}}}
\newcommand{\labSNR}{\labsty{\textbf{SNR}}}
\newcommand{\labM}{\labsty{M-index}}
\newcommand{\labENLL}{\labsty{ENL}}
\newcommand{\labH}{\labsty{\textbf{$\delta h$}}}
\newcommand{\labENL}{\labsty{\textbf{$r_{\widehat{ENL}}$}}}
\newcommand{\labMeanRatio}{\labsty{\textbf{$\mu_{N}$}}}
\newcommand{\labKL}{\labsty{\textbf{$D_{KL}$}}}
\newcommand{\resENL}{r_{\widehat{ENL}} }
\newcommand{\homo}{\delta h }
\newcommand{\meanratio}{\mu_{N} }
\newcommand{\resMeanRatio}{r_\mu }
\newcommand{\kl}{D_{KL} }

\newcommand{\csk}{CSK}
\newcommand{\tsx}{TSX}
\newcommand{\radarsat}{RADARSAT2}
\newcommand{\sent}{Sentinel-1}

\newcommand{\image}{\pgfuseimage}
\newcommand{\figpath}{./figures/}

\pgfdeclareimage[width=0.7\columnwidth]{detection}{\figpath detection2.pdf}

\pgfdeclareimage[width=0.2\columnwidth]{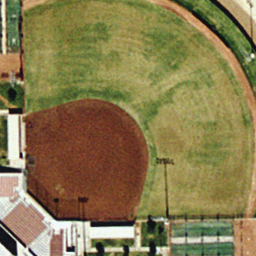}{\figpath data1.png}
\pgfdeclareimage[width=0.2\columnwidth]{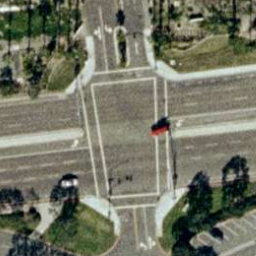}{\figpath data2.png}
\pgfdeclareimage[width=0.2\columnwidth]{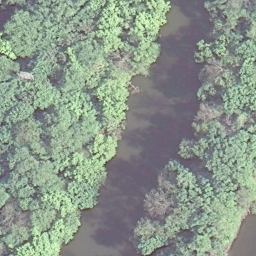}{\figpath data3.png}
\pgfdeclareimage[width=0.2\columnwidth]{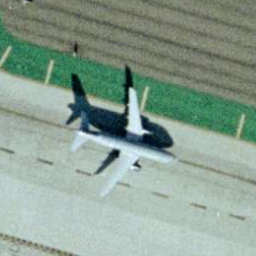}{\figpath data4.png}
\pgfdeclareimage[width=0.2\columnwidth]{ref}{\figpath 03000.png}
\pgfdeclareimage[width=0.2\columnwidth]{speckle}{\figpath f.png}
\pgfdeclareimage[width=0.2\columnwidth]{noisy}{\figpath 03000_1_speckled.png}

\pgfdeclareimage[width=0.8\textwidth]{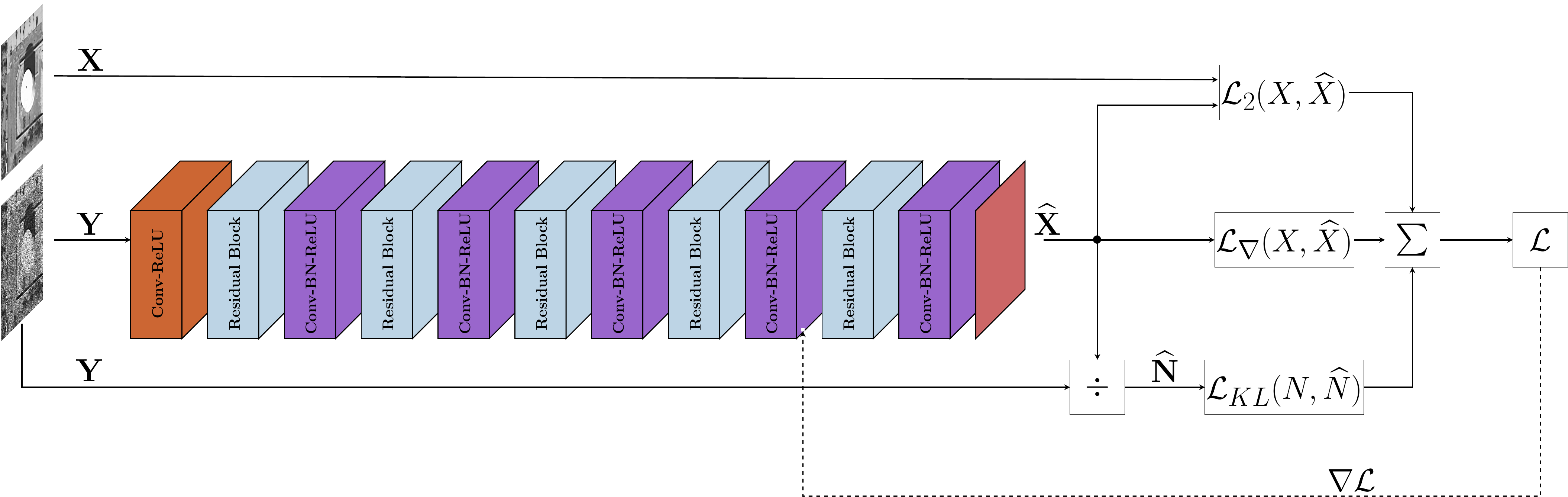}{\figpath net.pdf}
\pgfdeclareimage[width=0.2\textwidth]{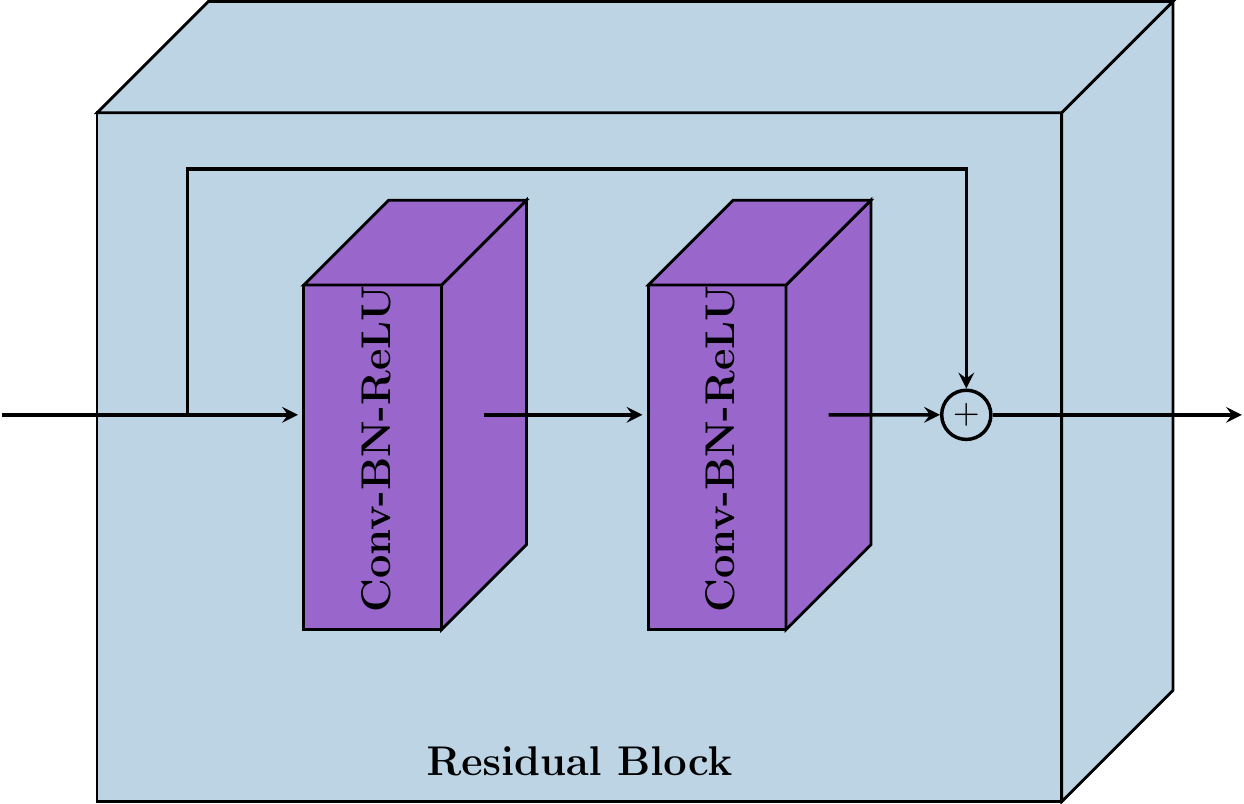}{\figpath res_block.pdf}

\pgfdeclareimage[width=0.5\columnwidth]{pavia_distr}{\figpath 	/pavia_g0.pdf}
\pgfdeclareimage[width=0.5\columnwidth]{sim_distr}{\figpath 	/sim.pdf}

%% file: imgs_sim.tex
\newcommand{\imsize}{0.11\textwidth}
\newcommand{\simpath}{./details/sim/}

\pgfdeclareimage[width=\imsize]{sim_noisy2}{\simpath 	/denseresidential_17_noisy.png}
\pgfdeclareimage[width=\imsize]{sim_ref2}{\simpath 		/denseresidential_17_ref.png}
\pgfdeclareimage[width=\imsize]{sim_prop2}{\simpath 	/denseresidential_17_prop_skip.png}
\pgfdeclareimage[width=\imsize]{sim_fans2}{\simpath 	/denseresidential_17_fans.png}
\pgfdeclareimage[width=\imsize]{sim_sarbm3d2}{\simpath 	/denseresidential_17_sarbm3d.png}
\pgfdeclareimage[width=\imsize]{sim_noland2}{\simpath 	/denseresidential_17_noland.png}
\pgfdeclareimage[width=\imsize]{sim_idcnn2}{\simpath 	/denseresidential_17_IDCNN.png}
\pgfdeclareimage[width=\imsize]{sim_sardrn2}{\simpath 	/denseresidential_17_SARDRN.png}

\pgfdeclareimage[width=\imsize]{sim_noisy3}{\simpath 	/storagetanks_6_noisy.png}
\pgfdeclareimage[width=\imsize]{sim_ref3}{\simpath 		/storagetanks_6_ref.png}
\pgfdeclareimage[width=\imsize]{sim_prop3}{\simpath 	/storagetanks_6_prop_skip.png}
\pgfdeclareimage[width=\imsize]{sim_fans3}{\simpath 	/storagetanks_6_fans.png}
\pgfdeclareimage[width=\imsize]{sim_sarbm3d3}{\simpath 	/storagetanks_6_sarbm3d.png}
\pgfdeclareimage[width=\imsize]{sim_noland3}{\simpath 	/storagetanks_6_noland.png}
\pgfdeclareimage[width=\imsize]{sim_idcnn3}{\simpath 	/storagetanks_6_IDCNN.png}
\pgfdeclareimage[width=\imsize]{sim_sardrn3}{\simpath 	/storagetanks_6_SARDRN.png}

\pgfdeclareimage[width=\imsize]{sim_noisy4}		{\simpath 	/denseresidential_14_noisy.png}
\pgfdeclareimage[width=\imsize]{sim_ref4}		{\simpath 	/denseresidential_14_ref.png}
\pgfdeclareimage[width=\imsize]{sim_prop4}		{\simpath 	/denseresidential_14_prop_skip.png}
\pgfdeclareimage[width=\imsize]{sim_fans4}		{\simpath 	/denseresidential_14_fans.png}
\pgfdeclareimage[width=\imsize]{sim_sarbm3d4}	{\simpath 	/denseresidential_14_sarbm3d.png}
\pgfdeclareimage[width=\imsize]{sim_noland4}	{\simpath 	/denseresidential_14_noland.png}
\pgfdeclareimage[width=\imsize]{sim_idcnn4}		{\simpath 	/denseresidential_14_IDCNN.png}
\pgfdeclareimage[width=\imsize]{sim_sardrn4}	{\simpath 	/denseresidential_14_SARDRN.png}

\pgfdeclareimage[width=\imsize]{sim_noisy5}{\simpath 	/storagetanks_8_noisy.png}
\pgfdeclareimage[width=\imsize]{sim_ref5}{\simpath 		/storagetanks_8_ref.png}
\pgfdeclareimage[width=\imsize]{sim_prop5}{\simpath 	/storagetanks_8_prop_skip.png}
\pgfdeclareimage[width=\imsize]{sim_fans5}{\simpath 	/storagetanks_8_fans.png}
\pgfdeclareimage[width=\imsize]{sim_sarbm3d5}{\simpath 	/storagetanks_8_sarbm3d.png}
\pgfdeclareimage[width=\imsize]{sim_noland5}{\simpath 	/storagetanks_8_noland.png}
\pgfdeclareimage[width=\imsize]{sim_idcnn5}{\simpath 	/storagetanks_8_IDCNN.png}
\pgfdeclareimage[width=\imsize]{sim_sardrn5}{\simpath 	/storagetanks_8_SARDRN.png}

%% file: imgs_real.tex
\newcommand{\imsizeReal}{0.14\textwidth}
\newcommand{\imsizeSub}{0.23\textwidth}
\newcommand{\imsizeAblation}{0.19\columnwidth}
\newcommand{\imsizeDetection}{0.19\columnwidth}
\newcommand{\realpath}{./details/real/}

\newcommand{\area}{/y_pavia/full/}
\pgfdeclareimage[width=\imsizeReal]{pavia_noisy}{\realpath 	\area  /noisy_enl.png}
\pgfdeclareimage[width=\imsizeReal]{pavia_prop}{\realpath 	\area  /prop_skip.png}
\pgfdeclareimage[width=\imsizeReal]{pavia_sarbm3d}{\realpath	\area  /sarbm3d.png}
\pgfdeclareimage[width=\imsizeReal]{pavia_noland}{\realpath 	\area  /noland.png}
\pgfdeclareimage[width=\imsizeReal]{pavia_fans}{\realpath 	\area  /fans.png}
\pgfdeclareimage[width=\imsizeReal]{pavia_idcnn}{\realpath 	\area  /IDCNN.png}
\pgfdeclareimage[width=\imsizeReal]{pavia_sardrn}{\realpath 	\area  /SARDRN.png}

\renewcommand{\area}{/y_pavia/(170,160)}
\pgfdeclareimage[width=\imsizeReal]{pavia_noisy_d1}{\realpath 	\area  /noisy_box.png}
\pgfdeclareimage[width=\imsizeReal]{pavia_prop_d1}{\realpath 	\area  /prop_skip.png}
\pgfdeclareimage[width=\imsizeReal]{pavia_sarbm3d_d1}{\realpath	\area  /sarbm3d.png}
\pgfdeclareimage[width=\imsizeReal]{pavia_noland_d1}{\realpath 	\area  /noland.png}
\pgfdeclareimage[width=\imsizeReal]{pavia_fans_d1}{\realpath 	\area  /fans.png}
\pgfdeclareimage[width=\imsizeReal]{pavia_idcnn_d1}{\realpath 	\area  /IDCNN.png}
\pgfdeclareimage[width=\imsizeReal]{pavia_sardrn_d1}{\realpath 	\area  /SARDRN.png}
\pgfdeclareimage[width=\imsizeSub]{pavia_subimg}{\realpath 	\area  /sub_img.pdf}

\pgfdeclareimage[width=\imsizeReal]{pavia_prop_ratio_d1}{\realpath 	\area  /ratioprop_skip.png}
\pgfdeclareimage[width=\imsizeReal]{pavia_sarbm3d_ratio_d1}{\realpath \area  /ratiosarbm3d.png}
\pgfdeclareimage[width=\imsizeReal]{pavia_noland_ratio_d1}{\realpath 	\area  /rationoland.png}
\pgfdeclareimage[width=\imsizeReal]{pavia_fans_ratio_d1}{\realpath 	\area  /ratiofans.png}
\pgfdeclareimage[width=\imsizeReal]{pavia_idcnn_ratio_d1}{\realpath 	\area  /ratioIDCNN.png}
\pgfdeclareimage[width=\imsizeReal]{pavia_sardrn_ratio_d1}{\realpath 	\area  /ratioSARDRN.png}

\renewcommand{\area}{/y_pavia/(860,650)}
\pgfdeclareimage[width=\imsizeAblation]{pavia_noisy_abl}{\realpath 	\area  /noisy.png}
\pgfdeclareimage[width=\imsizeAblation]{pavia_l2_abl}{\realpath 	\area  /l2.png}
\pgfdeclareimage[width=\imsizeAblation]{pavia_l2kl_abl}{\realpath 	\area  /l2kl.png}
\pgfdeclareimage[width=\imsizeAblation]{pavia_l2edge_abl}{\realpath 	\area  /l2edge.png}
\pgfdeclareimage[width=\imsizeAblation]{pavia_prop_abl}{\realpath 	\area  /prop_skip.png}

\renewcommand{\area}{/y_pavia/detection/full/}
\pgfdeclareimage[width=\imsizeReal]{pavia_noisy_det}{\realpath 	\area  /noisy.png}
\pgfdeclareimage[width=\imsizeReal]{pavia_prop_det}{\realpath 	\area  /prop_skip.png}
\pgfdeclareimage[width=\imsizeReal]{pavia_prop_ratio_det}{\realpath 	\area  /ratioprop_skip.png}
\pgfdeclareimage[width=\imsizeReal]{pavia_prop_mask}{\realpath 	\area  /mask_prop_skip.png}

\renewcommand{\area}{/y_pavia/detection/(877,662)/}
\pgfdeclareimage[width=\imsizeDetection]{pavia_noisy_det_d1}{\realpath 	\area  /noisy.png}
\pgfdeclareimage[width=\imsizeDetection]{pavia_prop_det_d1}{\realpath 	\area  /prop_skip.png}
\pgfdeclareimage[width=\imsizeDetection]{pavia_idcnn_det_d1}{\realpath 	\area  /IDCNN.png}
\pgfdeclareimage[width=\imsizeDetection]{pavia_sardrn_det_d1}{\realpath 	\area  /SARDRN.png}
\pgfdeclareimage[width=\imsizeDetection]{pavia_prop_ratio_det_d1}{\realpath 	\area  /ratioprop_skip.png}
\pgfdeclareimage[width=\imsizeDetection]{pavia_idcnn_ratio_det_d1}{\realpath 	\area  /ratioIDCNN.png}
\pgfdeclareimage[width=\imsizeDetection]{pavia_sardrn_ratio_det_d1}{\realpath 	\area  /ratioSARDRN.png}
\pgfdeclareimage[width=\imsizeDetection]{pavia_prop_mask_d1}{\realpath 	\area  /mask_prop_skip.png}
\pgfdeclareimage[width=\imsizeDetection]{pavia_idcnn_mask_d1}{\realpath 	\area  /mask_IDCNN.png}
\pgfdeclareimage[width=\imsizeDetection]{pavia_sardrn_mask_d1}{\realpath 	\area  /mask_SARDRN.png}

\renewcommand{\area}{/y_phoenix/full/}

\pgfdeclareimage[width=\imsizeReal]{phoenix_noisy}{\realpath 	\area  /noisy_enl.png}
\pgfdeclareimage[width=\imsizeReal]{phoenix_prop}{\realpath 	\area  /prop_skip.png}
\pgfdeclareimage[width=\imsizeReal]{phoenix_sarbm3d}{\realpath	\area  /sarbm3d.png}
\pgfdeclareimage[width=\imsizeReal]{phoenix_noland}{\realpath 	\area  /noland.png}
\pgfdeclareimage[width=\imsizeReal]{phoenix_fans}{\realpath 	\area  /fans.png}
\pgfdeclareimage[width=\imsizeReal]{phoenix_idcnn}{\realpath 	\area  /IDCNN.png}
\pgfdeclareimage[width=\imsizeReal]{phoenix_sardrn}{\realpath 	\area  /SARDRN.png}

\renewcommand{\area}{/y_phoenix/(793,244)}
\pgfdeclareimage[width=\imsizeReal]{phoenix_noisy_d1}{\realpath 	\area  /noisy_box.png}
\pgfdeclareimage[width=\imsizeReal]{phoenix_prop_d1}{\realpath 	\area  /prop_skip.png}
\pgfdeclareimage[width=\imsizeReal]{phoenix_sarbm3d_d1}{\realpath	\area  /sarbm3d.png}
\pgfdeclareimage[width=\imsizeReal]{phoenix_noland_d1}{\realpath 	\area  /noland.png}
\pgfdeclareimage[width=\imsizeReal]{phoenix_fans_d1}{\realpath 	\area  /fans.png}
\pgfdeclareimage[width=\imsizeReal]{phoenix_idcnn_d1}{\realpath 	\area  /IDCNN.png}
\pgfdeclareimage[width=\imsizeReal]{phoenix_sardrn_d1}{\realpath 	\area  /SARDRN.png}
\pgfdeclareimage[width=\imsizeSub]{phoenix_subimg}{\realpath 	\area  /sub_img.pdf}

\pgfdeclareimage[width=\imsizeReal]{phoenix_prop_ratio_d1}{\realpath 	\area  /ratioprop_skip.png}
\pgfdeclareimage[width=\imsizeReal]{phoenix_sarbm3d_ratio_d1}{\realpath \area  /ratiosarbm3d.png}
\pgfdeclareimage[width=\imsizeReal]{phoenix_noland_ratio_d1}{\realpath 	\area  /rationoland.png}
\pgfdeclareimage[width=\imsizeReal]{phoenix_fans_ratio_d1}{\realpath 	\area  /ratiofans.png}
\pgfdeclareimage[width=\imsizeReal]{phoenix_idcnn_ratio_d1}{\realpath 	\area  /ratioIDCNN.png}
\pgfdeclareimage[width=\imsizeReal]{phoenix_sardrn_ratio_d1}{\realpath 	\area  /ratioSARDRN.png}

\renewcommand{\area}{/y_phoenix/(550,430)}
\pgfdeclareimage[width=\imsizeAblation]{phoenix_noisy_abl}{\realpath 	\area  /noisy.png}
\pgfdeclareimage[width=\imsizeAblation]{phoenix_l2_abl}{\realpath 	\area  /l2.png}
\pgfdeclareimage[width=\imsizeAblation]{phoenix_l2kl_abl}{\realpath 	\area  /l2kl.png}
\pgfdeclareimage[width=\imsizeAblation]{phoenix_l2edge_abl}{\realpath 	\area  /l2edge.png}
\pgfdeclareimage[width=\imsizeAblation]{phoenix_prop_abl}{\realpath 	\area  /prop_skip.png}

\renewcommand{\area}{/y_tehran/full/}

\pgfdeclareimage[width=\imsizeReal]{tehran_noisy}{\realpath 	\area  /noisy_enl.png}
\pgfdeclareimage[width=\imsizeReal]{tehran_prop}{\realpath 	\area  /prop_skip.png}
\pgfdeclareimage[width=\imsizeReal]{tehran_sarbm3d}{\realpath	\area  /sarbm3d.png}
\pgfdeclareimage[width=\imsizeReal]{tehran_noland}{\realpath 	\area  /noland.png}
\pgfdeclareimage[width=\imsizeReal]{tehran_fans}{\realpath 	\area  /fans.png}
\pgfdeclareimage[width=\imsizeReal]{tehran_idcnn}{\realpath 	\area  /IDCNN.png}
\pgfdeclareimage[width=\imsizeReal]{tehran_sardrn}{\realpath 	\area  /SARDRN.png}

\renewcommand{\area}{/y_tehran/(16,127)}
\pgfdeclareimage[width=\imsizeReal]{tehran_noisy_d1}{\realpath 	\area  /noisy_box.png}
\pgfdeclareimage[width=\imsizeReal]{tehran_prop_d1}{\realpath 	\area  /prop_skip.png}
\pgfdeclareimage[width=\imsizeReal]{tehran_sarbm3d_d1}{\realpath	\area  /sarbm3d.png}
\pgfdeclareimage[width=\imsizeReal]{tehran_noland_d1}{\realpath 	\area  /noland.png}
\pgfdeclareimage[width=\imsizeReal]{tehran_fans_d1}{\realpath 	\area  /fans.png}
\pgfdeclareimage[width=\imsizeReal]{tehran_idcnn_d1}{\realpath 	\area  /IDCNN.png}
\pgfdeclareimage[width=\imsizeReal]{tehran_sardrn_d1}{\realpath 	\area  /SARDRN.png}
\pgfdeclareimage[width=\imsizeSub]{tehran_subimg}{\realpath 	\area  /sub_img.pdf}

\pgfdeclareimage[width=\imsizeReal]{tehran_prop_ratio_d1}{\realpath 	\area  /ratioprop_skip.png}
\pgfdeclareimage[width=\imsizeReal]{tehran_sarbm3d_ratio_d1}{\realpath \area  /ratiosarbm3d.png}
\pgfdeclareimage[width=\imsizeReal]{tehran_noland_ratio_d1}{\realpath 	\area  /rationoland.png}
\pgfdeclareimage[width=\imsizeReal]{tehran_fans_ratio_d1}{\realpath 	\area  /ratiofans.png}
\pgfdeclareimage[width=\imsizeReal]{tehran_idcnn_ratio_d1}{\realpath 	\area  /ratioIDCNN.png}
\pgfdeclareimage[width=\imsizeReal]{tehran_sardrn_ratio_d1}{\realpath 	\area  /ratioSARDRN.png}

\renewcommand{\area}{/y_tehran/(250,200)}
\pgfdeclareimage[width=\imsizeAblation]{tehran_noisy_abl}{\realpath 	\area  /noisy.png}
\pgfdeclareimage[width=\imsizeAblation]{tehran_l2_abl}{\realpath 	\area  /l2.png}
\pgfdeclareimage[width=\imsizeAblation]{tehran_l2kl_abl}{\realpath 	\area  /l2kl.png}
\pgfdeclareimage[width=\imsizeAblation]{tehran_l2edge_abl}{\realpath 	\area  /l2edge.png}
\pgfdeclareimage[width=\imsizeAblation]{tehran_prop_abl}{\realpath 	\area  /prop_skip.png}

\renewcommand{\area}{/y_accra2/full/}

\pgfdeclareimage[width=\imsizeReal]{accra2_noisy}{\realpath 	\area  /noisy_enl.png}
\pgfdeclareimage[width=\imsizeReal]{accra2_prop}{\realpath 	\area  /prop_skip.png}
\pgfdeclareimage[width=\imsizeReal]{accra2_sarbm3d}{\realpath	\area  /sarbm3d.png}
\pgfdeclareimage[width=\imsizeReal]{accra2_noland}{\realpath 	\area  /noland.png}
\pgfdeclareimage[width=\imsizeReal]{accra2_fans}{\realpath 	\area  /fans.png}
\pgfdeclareimage[width=\imsizeReal]{accra2_idcnn}{\realpath 	\area  /IDCNN.png}
\pgfdeclareimage[width=\imsizeReal]{accra2_sardrn}{\realpath 	\area  /SARDRN.png}

\renewcommand{\area}{/y_accra2/(300,770)}
\pgfdeclareimage[width=\imsizeReal]{accra2_noisy_d1}{\realpath 	\area  /noisy_box.png}
\pgfdeclareimage[width=\imsizeReal]{accra2_prop_d1}{\realpath 	\area  /prop_skip.png}
\pgfdeclareimage[width=\imsizeReal]{accra2_sarbm3d_d1}{\realpath	\area  /sarbm3d.png}
\pgfdeclareimage[width=\imsizeReal]{accra2_noland_d1}{\realpath 	\area  /noland.png}
\pgfdeclareimage[width=\imsizeReal]{accra2_fans_d1}{\realpath 	\area  /fans.png}
\pgfdeclareimage[width=\imsizeReal]{accra2_idcnn_d1}{\realpath 	\area  /IDCNN.png}
\pgfdeclareimage[width=\imsizeReal]{accra2_sardrn_d1}{\realpath 	\area  /SARDRN.png}
\pgfdeclareimage[width=\imsizeSub]{accra2_subimg}{\realpath 	\area  /sub_img.pdf}

\pgfdeclareimage[width=\imsizeReal]{accra2_prop_ratio_d1}{\realpath 	\area  /ratioprop_skip.png}
\pgfdeclareimage[width=\imsizeReal]{accra2_sarbm3d_ratio_d1}{\realpath \area  /ratiosarbm3d.png}
\pgfdeclareimage[width=\imsizeReal]{accra2_noland_ratio_d1}{\realpath 	\area  /rationoland.png}
\pgfdeclareimage[width=\imsizeReal]{accra2_fans_ratio_d1}{\realpath 	\area  /ratiofans.png}
\pgfdeclareimage[width=\imsizeReal]{accra2_idcnn_ratio_d1}{\realpath 	\area  /ratioIDCNN.png}
\pgfdeclareimage[width=\imsizeReal]{accra2_sardrn_ratio_d1}{\realpath 	\area  /ratioSARDRN.png}

%% file: 1.introduction.tex
\section{Introduction}

\IEEEPARstart{S}{ynthetic Apertur Radar}  (SAR) imaging system produces images affected by a multiplicative noise, called speckle, 
creating a succession of strong and weak backscatterings. 
The presence of the speckle impairs the performance of several tasks like detection, segmentation and classification, indeed a despeckling operation is crucial for the interpretation of SAR images.

The high number of studies and algorithms developed in the last forty years testifies the importance of this topic.
Despite the great understanding of the speckle and its characteristics, despeckling is still an open issue far from being solved.

The first solutions work in the spatial domain, such as \cite{Lee1980, Frost1982, Kuan1985, Touzi88, Lopes1990} and \cite{Kuan1987, Lopes1990bis}: the first ones are based on a minimum-mean-square-error (MMSE) while the second ones propose a maximum \textit{a posteriori} (MAP) filter. 
These methods produce intense smoothing for reducing speckle in homogeneous areas that can not be applied on the edges.

Since the early 1990s, despeckling techniques operating in a transformed domain have been proposed \cite{Guo1994, Franceschetti1995, Gagnon1997, Argenti2002}. Filters based on such approach  often operate an homomorphic transformation in order to work with additive noise. 
These solutions embody a strong spatial adaptability in order to better preserve edges, which is a crucial issue in SAR despeckling.

A new research line in the despeckling domain has been drawn by the non local methods, that have shown very effective performances in preserving details, while removing noise \cite{Deledalle2009}.
Such methods look for similar patches in the image and merge them in order to produce targets pixels. Usually, differently from the previous solutions, statistics of the speckle and of the SAR backscattering \cite{Goodman1984, Frery1997} are taken into account for the definition of patch similarity.  Several algorithms have been defined within the non local paradigm, mainly by differentiating the choice of the similarity criterion or the merging function. For example, the non local paradigm based on different SAR similarity distances is applied by the methods proposed in \cite{Coupe2008, Zhong2009}. Whereas, a ratio-based metric is used in \cite{Feng2011, Ferraioli2019bis}.
Hybrid approaches arose like \cite{Parrilli2012, Zhong2011,Cozzolino2014} that join the non local paradigm with the wavelet transform.
A detailed review of the aforementioned despeckling filters can be found in \cite{touzi2002,Deledalle2014}. 

In the last years, deep learning (DL) is showing great performance in many natural image processing tasks such as classification, detection, segmentation and not less denoising.
Indeed, also remote sensing community is starting to exploit the potential of this approach, even if many difficulties arise due to the difference among natural and remote sensed images.

Recently, several DL solutions have been proposed for SAR despeckling. Such methods are data driven: differently from the previous classical approaches, it is mandatory to have a dataset composed of many couples of noisy inputs and noise-free images (references). Since for SAR despeckling, a noise-free reference is not available, the first issue for such methods is  the construction of simulated dataset. 

Mainly, DL despeckling algorithms rely on the simulation of fully developed speckle multiplied to the gray scale version of an optical image, that at the same time serves as clean reference for the network. For sake of simplicity, this approach is referred as \textit{synthetic approach} in the following of the paper.
Among them we recall \cite{Wang2017, Zhang2018, Lattari2019, Yang2019, Denis2019}.
In \cite{Wang2017} a simple residual CNN composed of eight layers is proposed, while a CNN with dilated convolution in order to increase the receptive field and skip connections for avoiding vanishing gradient is presented in \cite{Zhang2018}. In \cite{Lattari2019} the use of U-Net has been proposed.
In \cite{Yang2019} the Mulog \cite{Deledalle2017} framework combined with an AWGN denoising CNN is adapted for SAR. Later, in \cite{Denis2019} the same method is proposed trying to combine DL and the NL paradigm trough a post classification of filtered image. 

Moreover, instead of using synthetic approach, in other techniques such as \cite{Chierchia2017,Cozzolino2020},
the multitemporal average version of SAR acquisition  serves as reference.
Always for sake of simplicity, this approach is referred as \textit{multitemporal approach} in the following of the paper.
Real data have been also used for training a CNN as in \cite{Ma2020} following the \textit{Noise2Noise} scheme \cite{Lehtinen2018}. In such scheme the network learns to predict the clean image by using as input-reference data two noisy images with same underlying clean data but different independent realizations of noise. 

Most of these proposal focus only on the definition of the architecture and use very similar cost functions not taking into account statistical properties of the SAR image  and the presence of strong scatterers, demanding their knowledge to the features extraction from the training data. 
In \cite{Zhang2018} and \cite{Ma2020} the mean-square-error (MSE) is used as cost function. In \cite{Wang2017} and \cite{Lattari2019} the MSE is combined with a total variation regularization. A smoothed $L_{1}$ loss adapted to the speckle noise case has been considered in \cite{Chierchia2017}. The first attempt to include first order statistics of the speckle was proposed in \cite{vitale2019}, whereas in \cite{Cozzolino2020} a cost function based on statistic similarity is used.

In this paper a CNN for SAR despeckling that takes into account statistical properties of the SAR image has been proposed. The network is a seventeen layer CNN with skip connection trained with the synthetic approach. Beyond the proposed architecture, the main contribution is in the definition of a multi-objective cost function given by combination of three terms, each designed for a precise goal. Indeed, each of this term takes care respectively of spatial details, statistical properties and strong scatterers identification.

The rest of the paper is organized as follows. The description of the method and related contribution is in Section \ref{sec:method}. Experimental results and discussion are presented in Section \ref{sec: discussion}. Conclusion are presented in Section \ref{sec:conclusion}. An ablation study of the cost function has been carried out in Appendix \ref{sec:appendix}.

%% file: 2.Method.tex
\section{Methodology}
\label{sec:method}
In this section the proposed method is described: first, the acquisition model and the statistics of the acquired SAR image is presented; then the definition of the data simulation process, of the proposed architecture and of the multi-objective cost function are detailed. Finally the contribution of the paper is highlighted.

\subsection{Signal Statistical Description}
\label{sec: speckle}
The interpretation of SAR image is challenging due to the geometrical properties of SAR imaging system and to the presence of speckle. 
Indeed, speckle is a multiplicative noise produced by interference among the backscatterings 
of the objects inside a resolution cell of the sensor \cite{Argenti2013}. 
The generic SAR image can be expressed like in Eq. (\ref{eq: sar_image})
\begin{equation}
	 Y = X \cdot N
	 \label{eq: sar_image}
\end{equation}
where $Y$ is the SAR image, $X$ the noise-free image and  $N$ the speckle.

The statistical distribution of the speckle is well known under certain conditions. Three main cases can be considered: homogeneous, heterogeneous and extremely heterogeneous areas. 
Homogeneous areas (such as fields, roads, etc...) are characterized by the lack of dominant scatterers and the surface $X$ can be considered stationary.
This is the case of the Fully Developed hypothesis for the speckle $N,$
whose amplitude follows the square root of Gamma distribution \cite{Frery1997}:
\begin{equation}
	p_N(n,L) = \frac{2L^L}{\Gamma(L)} n^{2L-1}e^{-Ln^2} \hspace{1cm} n,L>0
	\label{eq:speckle_distribution}
\end{equation}

where $L$ is the number of looks of the SAR image and $\Gamma(\cdot)$ is the Gamma function. This probability density function (pdf) in case of single look becomes a Rayleigh distribution. 

Heterogeneous (tree and forest) and extremely heterogeneous areas (urban), are characterized by objects with shape and dimension that produce geometrical distortions and strong backscattering (e.g. multiple bounces, layover and shadowing). 
In heterogeneous areas, the speckle can be still considered Gamma distributed but the surface is not stationary anymore.
In extremely heterogeneous area, the hypothesis of distributed scatterers is not valid anymore due to the presence of dominant ones. Indeed, the speckle does not follow anymore the fully developed hypothesis \cite{Tison2004}.

The statistical distribution of SAR backscattering $Y$ in different scenarios is provided in \cite{Frery1997} where the use of the square root of generalized inverse Gaussian distribution $\mathcal{G}_A(\alpha,\gamma,\lambda,L)$ as general model for the amplitude return of SAR backscattering is proposed. The authors prove that the distribution of the SAR return of homogeneous (HO), heterogeneous (H) and extremely heterogeneous (EH) areas are particular case of this distribution depending on the parameter subspace.
An extension of this classification considering several possible scenarios has been recently proposed in \cite{Yue2020}.

According to \cite{Frery1997}, in the subspace $(\alpha >0, \gamma =0 , \lambda>0,L>0)$, when $\alpha$ and $\lambda$ tends to infinite, the distribution tends to a square root gamma $\Gamma^{1/2}(L,L/\beta)$ as in Eq. (\ref{eq:speckle_distribution}) with $\beta$ being the estimation $E[Y^2]$ of the second order statistic. Such distribution describes the return from HO areas.

Moreover, the authors have proved that the SAR return $Y$ in H area follows the $\mathcal{K}_A(\alpha,\lambda,L)$ distribution. This is the distribution the $\mathcal{G}_A(\alpha,\gamma,\lambda,L)$ tends to, when the parameter subspace is always $(\alpha >0, \gamma =0 , \lambda>0,L>0).$

For EH areas the amplitude distribution of the SAR image can be described according to $G_A^0(\alpha,\gamma,L)$, that is the distribution the $\mathcal{G}_A(\alpha,\gamma,\lambda,L)$  tends to, when the parameter subspace is $(\alpha <0, \gamma >0 , \lambda=0,L>0).$

\subsection{Data Simulation}
\label{sec: data_simulation}
In this section the data simulation process adopted for the training of the proposed CNN based despeckling algorithm is illustrated.

\begin{figure}[h]
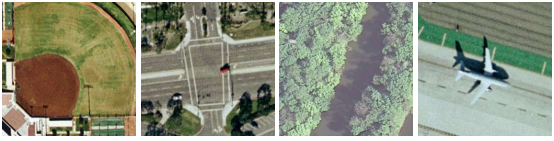

	\centering
	\begin{tabular}{cccc}
		\image{data1} & \image{data2} & \image{data3} & \image{data4} \\
	\end{tabular}
	\caption{RGB samples of Merced Land Use dataset}
	\label{fig: merced samples}
\end{figure}

Thousands of noise-free images from the optical UC Merced Land Use dataset \cite{MercedLandUse} have been considered. This dataset is typically considered for classification purposes thanks to the presence hundreds images belonging to different classes. Samples of this dataset are shown in  Fig. \ref{fig: merced samples}.

\begin{figure}[h]
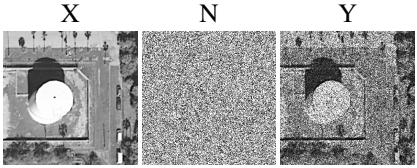

	\centering
	\begin{tabular}{ccc}
		X & N & Y\\
		\image{ref} & \image{speckle} & \image{noisy} \\
	\end{tabular}
	
	\caption{Simulation process, from left to right: noise-free reference, simulated noise, simulated SAR image}
	\label{fig: sim samples}
\end{figure}

The optical images have been converted from the RGB domain to the gray scale one obtaining the noise-free references $X$.
The speckle noise $N$ has been generated under the fully developed hypothesis in case of single look image according to Eq. (\ref{eq:speckle_distribution}). The final noisy image $Y$ has been obtained by simply multiplying the noise-free image by the speckle, as in Fig. \ref{fig: sim samples}.

\begin{figure}[ht]
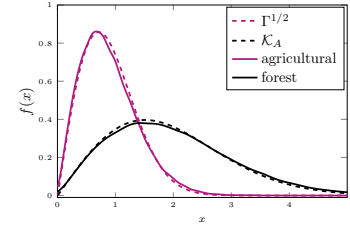

    \centering
    \image{sim_distr}
    \caption{Comparison between two different samples of training dataset: agricultural sample (magenta solid) and forest (black solid). In dashed the theoretical $\Gamma^{1/2}$ and $\mathcal{K}_A$ distributions}
    \label{fig:simulated-distribution}
\end{figure}

Considering that the Merced Land Use dataset is composed of several scenarios (such as agricultural field, baseball diamonds, forest, residential areas etc.), the simulation process transforms all these data in noisy images whose distribution belongs to  the $\Gamma^{1/2}$ distribution (HO areas) or to $\mathcal{K}_A$ distribution (H areas), that both are particular case of the generalized inverse Gaussian distribution.
In Fig. \ref{fig:simulated-distribution} the distributions of two samples of the dataset are depicted. The magenta solid curve represents the distribution of a simulated image taken from the  "agricultural"  class of the dataset. In this case the surface $X$ is almost homogeneous and the distribution of the resulting simulated $Y$ fits the $\Gamma^{1/2}$ (magenta dashed).
At the same time, the black solid curve represents the distribution of a simulated image taken from the "forest" class. In this case, the texture $X$ can not be considered homogeneous but some fluctuation had to be taken into account. Indeed, the distribution fits well the $\mathcal{K}_A$ curve. 
This process does not allow to simulate the EH case, where the speckle is not fully developed mainly due to presence of dominant scatterers and geometrical distortions. 

From the whole dataset, $57526\times64\times64$ amplitude patches for the training and $14336\times64\times64$ for the validation have been extracted.

\subsection{Network Architecture}
\begin{figure*}[t]
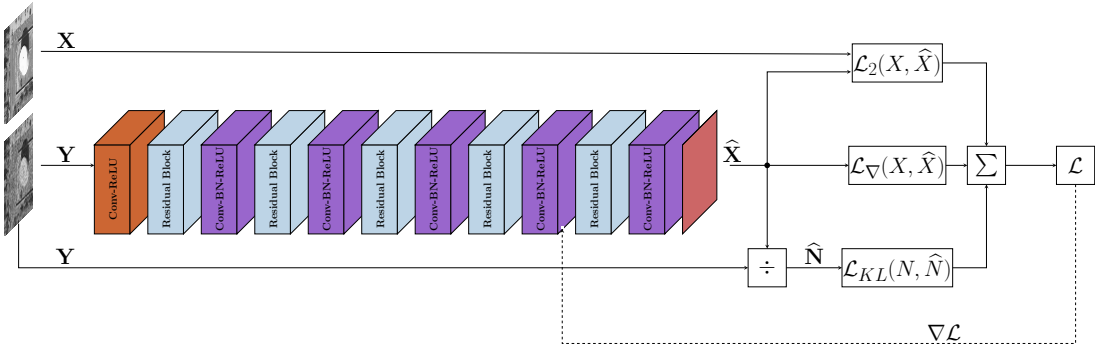

	\centering
	\image{net}
	\caption{Network architecture: all the layers have 64 features maps with $3\times3$ convolutional kernel. The first layer (in orange) is followed by ReLU activation function. After there is an alternation of residual block (in light blue) and inner layers with ReLU and batch normalization (in purple), while the last layer (in red) does not have neither activation function nor normalization. The cost function is a linear combination of three terms.}
	\label{fig: network}
\end{figure*}

\begin{figure}
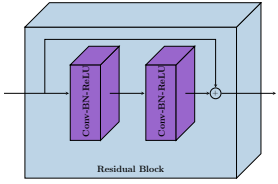

	\centering
	\image{res_block}
	\caption{Residual Block is composed of two Conv-BN-ReLU layers and a skip connection that sums the input to the output of second layer.}
	\label{fig: residual block}
\end{figure}
The design of the proposed network architecture comes from the results achieved in our previous works \cite{vitale2019} and \cite{vitale2020}, where ten layers CNN with different cost function were proposed.

Starting from the result of  \cite{vitale2020}, the proposed neural network is composed of seventeen convolutional layers. For each layer we consider ReLU as activation function \cite{Krizhevsky2012}, but for the last.
In all the layers batch normalization \cite{Ioffe2015} is performed except for the first and the last ones.
In addition, skip connections, that have shown great utility in training deep networks \cite{He2015}, are introduced in the inner layers.

Given the previous remarks, the output of layer $k$ can be expressed as:

\begin{equation}
z_k=f_k(\Phi_k,z_{k-1}) =  \nonumber
\end{equation}
\small

\begin{equation}
\begin{cases}
\sigma \left( w_k \ast Y + b_k \right) & \text{k=1} \\
\\
BN\left[\sigma \left( w_k \ast z_{k-1} + b_k \right)\right] +\alpha f_{k-3}  (\Phi_{k-3},z_{k-4})   & \text{1$<$k$<$D} \\
\\
\left( w_k \ast z_{k-1} + b_k \right) & \text{k=D}  \\
\end{cases}
\end{equation}

\normalsize
where 
$$
\alpha =
\begin{cases}
1 & \left< k-1 \right>_3=0 \\
0 & otherwise \\
\end{cases}
$$

with number of layers D=17, and $( w_k, b_k,\Phi_k,z_k)$ the weights, the bias, the set of parameters and the output of layer $k$, respectively. $BN$ stays for batch normalization and $\sigma(\cdot) = max(0,\cdot)$ is the ReLU activation function.
The operation $\left< k-1 \right>_3$ is the reminder of the division $(k-1)/3$.

Based on this network architecture, given a couple of samples $(Y,X)$ where $Y$ is the noisy image and $X$ acts as reference, the final estimated clean image is $\hx=z_D$.

For each layer 64 features maps are extracted except for the last one that has to fit the single channel output.
All the convolutional kernel have dimension $3 \times 3$. 
In Fig.\ref{fig: network}, a scheme of the network is depicted. The scheme of residual block is defined in Fig. \ref{fig: residual block}.

\subsection{Cost Function}

In the proposed algorithm, the aim is to propose a cost function that takes care both of spatial and statistical properties of the SAR images.
The defined multi-objective cost function $\mathcal{L}$ is a linear combination of three terms, each of them specifically dedicated to catch and to preserve information from the SAR image. Specifically:

\begin{equation} \label{eq: cost}
\mathcal{L} =  \mathcal{L}_{2} + \lambda_{KL} \mathcal{L}_{KL} + \lambda_\nabla \mathcal{L}_{\nabla}
\end{equation}

\begin{equation}
\mathcal{L}_{2} =   \mathcal{L}_{MSE} = || \hx - X ||^2 \end{equation} 
\begin{equation}
    \mathcal{L}_{KL} =  D_{KL} (  \hn, N_{teo}   )  
\end{equation}
\begin{equation}
    \mathcal{L}_{\nabla} = \left|\left| \nabla {X} - \nabla {\hx}\right|\right| ^2 
\end{equation}

$\lmse$ is the mean square error
between the  reference $X$ and filtered image $\hx$;  
$\lkl$ is the Kullback-Leibler divergence ($D_{KL}$) between the distribution of estimated noise $\hn = Y/\hx$ and the theoretical one $N_{teo}$, whose definition is provided in section \ref{sec: metrics}; 
$\ledge$ is the MSE between the gradient of the reference $X$ and gradient of the filtered image $\hx$.

Let us consider each of the three terms separately.

Naturally, the goal is to train the network to generate an output as similar as possible to the reference. 
To this aim the $\lmse$ term directly compares the output $\hx$ with the reference $X$ and it is responsible of spatial reconstruction.

Despite the importance of reducing spatial distortion, taking into account the properties of the noise within the despeckling operation is crucial, as shown by different methods like \cite{Feng2011, Ferraioli2019bis}. For this reason, the $\lkl$ term that takes into account the statistical properties of the noise has been introduced.

The $\lkl$ is the Kullback-Leibler divergence computed between the pdf of the estimated ratio image (the ratio between the SAR image and the estimated noise free one) and the theoretical fully developed speckle (in our case a Rayleigh distribution with parameter $\sigma=1/\sqrt{2}$).
The goal is to train the network to produce an output whose ratio image follows the statistical properties of the speckle.

The introduction of the $\ledge$ term is two fold: improving the edge preservation \cite{Xu2015} and dealing with dominant scatterers in real images. $\ledge$ compares the gradients of $\hx$ with the gradient of $X$.  The gradient gives information on the edges but, obviously, is not exactly an edge detector. It highlights transitions in images and so tends to identify the presence of structures.
So, if from one side it trains the network in preserving edges, on the other it helps the network in identifying and isolating strong scatterers.

In Appendix \ref{sec:appendix} an ablation study on the effects of these three terms has been proposed.

\subsection{Identification of Not Fully Developed Areas}
\label{sec:detection}
The presence of strong scatterers is challenging for all the filters and their filtering policy is still an open issue within the despeckling community \cite{Argenti2013}.

As reported in literature, these points should be left unfiltered or at least processed in a different way \cite{Arienzo2020}.
Some methods, such as SAR-BM3D, NOLAND and FANS, filter them by aggregation of similar patches selected trough a statistical approach; other methods, such as \cite{Vitale2019bis}, do not filter them at all.

These points are related to EH areas (usually urban) where,  as pointed out by \textit{Frery et al} in \cite{Frery1997} and \textit{Tison et al.} in \cite{Tison2004}, the speckle is not fully developed anymore.

As reported in section \ref{sec: data_simulation}, the only distribution not included in the training is the one for the EH areas. 
However, the defined cost function allows an easy detection of such areas. Indeed, if from one side the presence of $\lkl$ encourages the filtering under the fully developed hypothesis, from the other side the $\ledge$ tends to preserve edges and to identify structures. Their combination (together with $\lmse$) highlights the presence of such points producing strong values on the ratio image.
As a matter of fact, on these points the ratio image of an ideal filter should not show a Rayleigh distribution.
Thus, the appearance of such points on the ratio image can be considered as a positive issue. Actually, it allows to identify such points, having a different statistical distribution from the trained one (i.e. it allows to automatically identify points belonging to EH areas).

The identification of such EH points is performed directly from the ratio image produced by the proposed CNN. For this goal, a combination of the ratio edge detector proposed in \cite{Touzi88} and a Kolomogorov-Smirnov test on the ratio image produced by our algorithm is applied. 

\begin{figure}
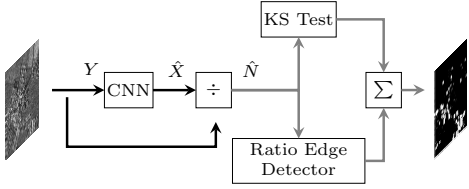

	\centering
	\image{detection}
	\caption{Flowchart for the identification of not fully developed areas from the ratio image.}
	\label{fig: detection chart}
\end{figure}

The former aims to highlight the edges and remaining structures in the ratio image, 
the latter detects the area where the predicted speckle is not fully developed by comparison through a threshold at patch level of ratio image distribution and the fully developed one. 
In Fig. \ref{fig: detection chart}, a flowchart of this detection process is depicted.
The knowledge of those pixels allows the final user to decide which filtering policy to be adopted (left unfiltered, define a specific statistical based filter,  using a multitemporal approach, etc…) \cite{Argenti2002}.

\subsection{Contribution}
In this section, the contribution of the proposed method, called  MONet (Multi-Objective Network for SAR despeckling), is described and its innovative issues are highlighted.
The proposed MONet shares some points with IDCNN \cite{Wang2017}, SAR-CNN \cite{Chierchia2017} and SAR-DRN\cite{Zhang2018}.
Indeed, the proposed CNN has seventeen layers like SAR-CNN, and also skip connections are added in the inner layers, like in SAR-DRN. Differently from SAR-DRN, a deeper network has been preferred to dilated convolutions. Deeper networks allow to extract more features and to add more abstractions, facilitating the exploitation of the data and the network generalization.The depth has been set experimentally: in \cite{vitale2020bis} it has been proved that deeper network gives better results.

The main innovation consists in the definition of the cost function: a combination of the $\lmse$ norm  with other terms is used for the reconstruction. 
While IDCNN combines the $\lmse$ with the total variation in order to provide smooth results, in the proposal the term $\ledge$ for edges preservation and dominant scatterers identification has been considered. Moreover, a statistical term $\lkl$ for speckle properties preservation  is added in the combination that leads to the whole cost function. 
In Tab.\ref{tab:diff_architectures} the differences among the aforementioned methods are summarised.
\begin{table}[h]
    \centering
    \caption{Main differences among compared DL methods}
    \label{tab:diff_architectures}
    \setlength{\tabcolsep}{4pt}
    \renewcommand\arraystretch{1.3}
    \begin{tabular}{|c|c|c|c|c|}
    \hline
        \labsty{Method} & \labsty{Depth} & \labsty{Skip Connection}&  \labsty{Spatial Loss} & \labsty{Statistical Loss}   \\
        \hline
        \labIdcnn       & 10             &   \xmark                    &     $\lmse + TV$       &        \xmark                   \\
        \labSarcnn      & 17             &  \xmark                      &  smoothed $\mathcal{L}_1$&    \xmark\\
        \labSardrn      & 7             &   \cmark                     &   $\lmse$                 & \xmark\\
        \labProp        & 17            & \cmark                      &  $\lmse + \ledge$          & $\lkl$ \\
        \hline
        
    \end{tabular}
    
\end{table}

%% file: 3.discussion.tex
\section{Experimental Results}
\label{sec: discussion}
In order to validate the method, experiments have been carried out on both simulated and real data. Both quantitative analysis, based on performance indexes, and qualitatively analysis, based on visual inspection,  have been conducted.

For comparison, two different families of despeckling algorithms have been considered: Non Local  and Deep Learning based ones. In particular, NL algorithms have been addressed since they are often considered in literature as a benchmark for evaluating achievable performances. Between the available NL algorithms we considered FANS \cite{Cozzolino2014}, SAR-BM3D \cite{Parrilli2012} and NOLAND \cite{Ferraioli2019bis}. 
While the DL based algorithms have been considered in order to compare the performances of the proposed algorithm with methods sharing the same philosophy. In particular, ID-CNN and SAR-DRN as deep learning methods have been used.

For each NL method, the parameters have been set accordingly to those suggested in the relative papers.
While, given that the DL solutions are data driven, in order to have a fair comparison 
the CNN based solutions have been re-trained on our same dataset following the description of the authors.

For this reason, we did not compare with the solution based on the multitemporal approach \cite{Chierchia2017}, \cite{Cozzolino2020}, \cite{Ma2020}, because a fair comparison is not possible using a training  on simulated data.
Moreover, for the rest of DL papers, the authors did not make available either their code or training dataset.

The proposed network is trained with mini batch of 128 samples, using the Adam optimizer \cite{Kingma14} with parameter $\beta_1=0.9$ and $\beta_2=0.99$. 
The learning rate is set to $\eta=0.0001$ for the first 87 epochs, and after the training is refined for other 35 epochs with a learning rate scaled by 10.
The lambdas parameter for the cost function have been  empirically set for balancing their effects: $\lambda_{KL}=10^4$ and $\lambda_\nabla = 1$. The framework used for the implementation is Theano, running on Python. Both training and testing have been carried out on a GeForce GTX 1080Ti GPU with 11 GB of memory. The code of proposed method is available in \textit{https://github.com/sergiovitale/MONet-SAR-Despceckling-CNN-Theano-implementation}

\begin{figure*}[t]
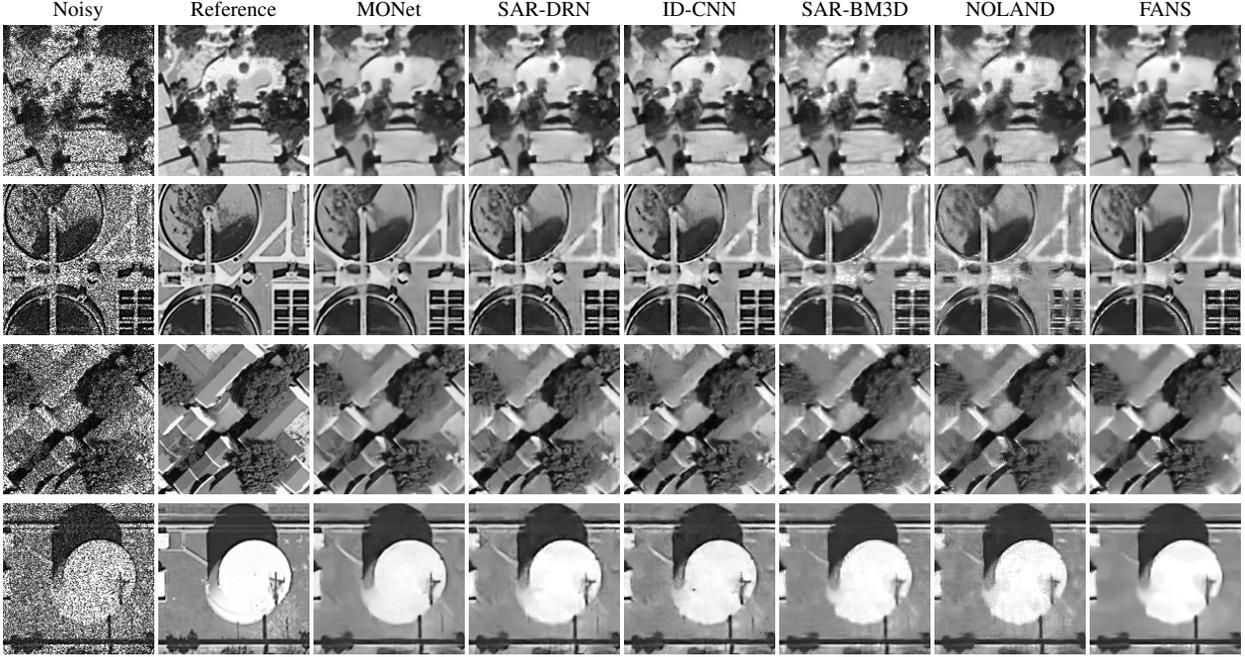

	\centering
	\begin{tabular}{cccccccc}
		\labNoise & \labRef & \labProp & \labSardrn& \labIdcnn & \labSarbm & \labNoland & \labFans \\
	\image{sim_noisy2} & \image{sim_ref2} & \image{sim_prop2} & \image{sim_sardrn2} & \image{sim_idcnn2} & \image{sim_sarbm3d2} & \image{sim_noland2} & \image{sim_fans2}\\
	\image{sim_noisy3} & \image{sim_ref3} & \image{sim_prop3} & \image{sim_sardrn3} & \image{sim_idcnn3} & \image{sim_sarbm3d3} & \image{sim_noland3} & \image{sim_fans3}\\
	\image{sim_noisy4} & \image{sim_ref4} & \image{sim_prop4} & \image{sim_sardrn4} & \image{sim_idcnn4} & \image{sim_sarbm3d4} & \image{sim_noland4} & \image{sim_fans4}\\
	\image{sim_noisy5} & \image{sim_ref5} & \image{sim_prop5} & \image{sim_sardrn5} & \image{sim_idcnn5} & \image{sim_sarbm3d5} & \image{sim_noland5} & \image{sim_fans5}\\
  
	\end{tabular}
	\caption{Results on a subset of the simulated images, from left to right: simulated noisy image, noise-free reference, MONet, SAR-DRN, ID-CNN, SARBM3D, NOLAND, FANS }
	\label{fig: simulated results}
\end{figure*}

\subsection{Metrics}
\label{sec: metrics}
For numerical evaluation both reference and no-reference metrics have been considered.
As reference metrics the \textit{Structural Similarity} (SSIM) index, the \textit{Mean Squared Error} (MSE) and the \textit{Signal to Noise Ratio} (SNR) have been used for evaluating results on the simulated dataset, where a reference is available.

\begin{itemize}
\item \textbf{SSIM} measures the similarity between  $\hx$ and $X$  from a perceptual point of view. The ideal filter would produce SSIM=1
\item \textbf{MSE} measures the average similarity between  $\hx$ and $X$ . The ideal value is zero.
\item \textbf{SNR} measures the signal to noise ratio and give us information about the capability of the noise suppression. The higher SNR, better  the filter.
\end{itemize}

Regarding no reference metrics the Equivalent Number of Looks (ENL),the M-index, the Haralick homogeneity $\homo$, the residual ENL $\resENL$, the mean of the ratio $\meanratio$ and the Kullback-Leibler divergence $\kl$ are considered.
\begin{itemize}
    \item \textbf{ENL}  is an indicator of noise suppression in homogeneous areas. Once a homogeneous area has been selected, the ENL computes the ratio between the squared power of the mean and the variance of the filtered image (both in intensity format).
		\begin{equation}
		    ENL = \frac{E[\hx^2]^2}{Var(\hx^2)}
		\end{equation}
		Higher is the ENL, greater is the noise suppression.\\
		
	\item \textbf{M-index}\cite{Gomez2017} is a combination of three factors $\homo$, $\resENL$ and $\resMeanRatio$:
	\begin{itemize}
		\item $\homo$ is based on the Haralick homogeneity texture \cite{Haralick1973} and it is the distance between the homogeneity  $h_0$ of ratio image compared with the homogeneity $h_g$ of the random permuted the ratio image itself.
		It is compute as $ \delta h = \frac{|h_0 - h_g |}{h_0} $, with  
	    \begin{equation}
	     h_z  = \sum_i \sum_j \frac{1}{1+(i-j)^2}\cdot p_z(i,j)
	     \end{equation}
		where $p_z(i,j)$ is the gray scale level co-occurrence matrix of the ratio image $z$ at an arbitrary position.
		$\homo$ computes a sort of correlation of the ratio image and give us information of remaining structures that should not be present after an ideal filtering.
		The ideal filter will produce $\homo=0$.

		\item $\resENL$ is the residual ENL and once \textit{n} homogeneous patches are selected the ENL computed on ratio and SAR image are compared.
		\begin{equation}
		    \labENL = \frac{1}{n} \sum_{i=1}^{n} \frac{|\widehat{ENL}_{noisy}(i)-\widehat{ENL}_{ratio}(i)|}  {\widehat{ENL}_{noisy}(i)}
		\end{equation} 	
		The ideal filter will produce $\resENL$ equal to 0.
		
		\item $\resMeanRatio$ is the function of the mean ratio $\meanratio$ computed on the same patches selected for the $\resENL$
		\begin{equation}
		    r_{\mu} =\frac{1}{n} \sum_{i=1}^{n} | 1- \mu_N(i)|
		\end{equation}
		
		The ideal filter will produce $\resMeanRatio$ equal to zero.

	\end{itemize}
	
	The ideal filter will produce an M-index equal to zero.
	
	\item the $\mathbf{\kl}$  computes the distance between the statistical distribution of the ratio image with the theoretical Rayleigh distribution.
	\begin{equation}
	     \kl(\hat{N},N_{teo})  = \sum_i P_{\hat{N}}(i) log_2 \left( \frac{P_{\hat{N}}(i)}{P_{N_{teo}}(i)} \right)
	\end{equation}
	where $P_{\hat{N}}$ is the pdf of the predicted speckle and $P_{N_{teo}}$ is the pdf of the theoretical noise.
	Under the fully developed hypothesis, an ideal filter will produce a $\kl = 0$
	
\end{itemize}
Clearly, other indexes could have been adopted and considered. We focus on these ones since they are largely and commonly adopted by the community. 

\subsection{Simulated Results}

\label{sec: simulated results}
For the simulation 100 single look amplitude images of size $256 \times 256$ have been selected. These belong to 5 classes (20 for each class) of the Merced Land Use dataset not used during the training phase.
In Tab. \ref{tab: num_ass}, the numerical evaluation for reference metrics and ENL, averaged on the whole dataset, is shown. Best solution is expressed in bold, the second best is underlined.

Regarding the reference metrics SSIM, SNR and MSE, it is evident that DL solutions outperform the other given that they are trained on a dataset with same properties of the testing one.
The proposed MONet outperforms all the DL and NL methods.
The best NL solution on simulated experiments is SAR-BM3D. 
Regarding the ENL, FANS performs largely better than the others, followed by the MONet.

Numerical assessment is not enough and visual inspection is essential for understanding the performance of a filter.
Four different images, with different textures are shown in Fig.\ref{fig: simulated results} for a qualitative analysis. Together with the noisy images (first column), the noise free reference images are reported. Columns from 3 to 8 show the filtering results of the different considered approaches.

Among the NL methods, FANS is over smoothed losing many spatial details, but with a good edge preservation.
NOLAND and SAR-BM3D are very close each other with a good detail preservation but both of them produce some artefacts on homogeneous areas that impair the edges preservation.
Among the CNN methods, MONet shows the best performance on spatial details and edges preservation. IDCNN and SAR-DRN are very close each other with the former producing a filtered image still a bit noisy and the latter producing some distortions on the edges.

Generally, the proposed solution seems to produce the most similar image to the reference, showing a very good noise suppression without losing details and a good edges preservation.

Regarding the computational efficiency, the processing time is clearly related to the number of parameters the network is composed of. Being our network deeper compared to the others, it allows to extract more representative features, resulting in a better generalization at the cost of higher computational time. Anyway, the algorithm guarantees a fast processing time: for example a 3000x3000 is processed in approximately 4 seconds.

\begin{table}[ht]
	\centering
	\caption{Numerical Assessment on Simulated Dataset: the value are averaged on the whole simulated testing dataset composed of 100 images}
	\setlength{\tabcolsep}{5pt}
	\renewcommand\arraystretch{1.3}
	\begin{tabular} {|l|cccc|}
		\hline
					& 	\labSSIM & \labSNR & \labMSE & \labENLL \\
		\hline
		 \labFans	&	.7049	&8.0432	&.00482	&\first{822}\\
		 \labSarbm	&	.7379	&8.4251	&.00438	&240\\
		 \labNoland	&	.6847	&7.4712	&.00544 &84\\
		 \hline                           
		 \labIdcnn	&	.7231	&8.3644	&.00437	&144\\  
		 \labSardrn	&	\second{.7437}	&\second{8.7240}	&\second{.00406}&374\\ 
		 \labProp	&	\first{.7510}	&\first{8.8555}	&\first{.00395}	&\second{580}\\
		 \hline
	\end{tabular}
	
	\label{tab: num_ass}
\end{table}

\subsection{Result on Real SAR Images}
\begin{figure*}
	\centering
	\begin{tabular}{ccccccc}
		\labNoise & \labProp & \labSardrn &\labIdcnn & \labNoland & \labSarbm & \labFans\\
		\image{pavia_noisy} 		& \image{pavia_prop} 			& \image{pavia_sardrn} 			& \image{pavia_idcnn} 			& \image{pavia_noland} 			& \image{pavia_sarbm3d}			& \image{pavia_fans}\\		
		\image{pavia_noisy_d1} 	& \image{pavia_prop_d1} 			& \image{pavia_sardrn_d1} 		& \image{pavia_idcnn_d1} 		& \image{pavia_noland_d1}		& \image{pavia_sarbm3d_d1}		& \image{pavia_fans_d1}\\
		& \image{pavia_prop_ratio_d1} 	& \image{pavia_sardrn_ratio_d1} 	& \image{pavia_idcnn_ratio_d1} 	& \image{pavia_noland_ratio_d1}	& \image{pavia_sarbm3d_ratio_d1}	& \image{pavia_fans_ratio_d1}\\
		
	\end{tabular}
	\caption{Results on \csk{} image: scene under test in the top row; details of the image in the second row; corresponding ratio image in the third row.}
	\label{fig:results_pavia}
\end{figure*}

\begin{figure*}
	\centering
	\begin{tabular}{ccccccc}
		\labNoise & \labProp & \labSardrn &\labIdcnn & \labNoland & \labSarbm & \labFans\\
		\image{phoenix_noisy} 		& \image{phoenix_prop} 			& \image{phoenix_sardrn} 			& \image{phoenix_idcnn} 			& \image{phoenix_noland} 			& \image{phoenix_sarbm3d}			& \image{phoenix_fans}\\		
		\image{phoenix_noisy_d1} 	& \image{phoenix_prop_d1} 			& \image{phoenix_sardrn_d1} 		& \image{phoenix_idcnn_d1} 		& \image{phoenix_noland_d1}		& \image{phoenix_sarbm3d_d1}		& \image{phoenix_fans_d1}\\
		& \image{phoenix_prop_ratio_d1} 	& \image{phoenix_sardrn_ratio_d1} 	& \image{phoenix_idcnn_ratio_d1} 	& \image{phoenix_noland_ratio_d1}	& \image{phoenix_sarbm3d_ratio_d1}	& \image{phoenix_fans_ratio_d1}\\

	\end{tabular}
	\caption{Results on \radarsat{} image:  scene under test in the top row; details of the image in the second row; corresponding ratio image in the third row.}
	\label{fig:results_phoenix}
\end{figure*}
\begin{figure*}[ht]
	\centering
	\begin{tabular}{ccccccc}
		\labNoise & \labProp & \labSardrn &\labIdcnn & \labNoland & \labSarbm & \labFans\\
		\image{tehran_noisy} 	 & \image{tehran_prop} 			& \image{tehran_sardrn} 			& \image{tehran_idcnn} 			& \image{tehran_noland} 			& \image{tehran_sarbm3d}			& \image{tehran_fans}\\
		\image{tehran_noisy_d1} & \image{tehran_prop_d1} 			& \image{tehran_sardrn_d1} 		& \image{tehran_idcnn_d1} 			& \image{tehran_noland_d1}			& \image{tehran_sarbm3d_d1}		& \image{tehran_fans_d1}\\
		 & \image{tehran_prop_ratio_d1} 	& \image{tehran_sardrn_ratio_d1} 	& \image{tehran_idcnn_ratio_d1} 	& \image{tehran_noland_ratio_d1}	& \image{tehran_sarbm3d_ratio_d1}	& \image{tehran_fans_ratio_d1}\\
	\end{tabular}
	\caption{Results on \tsx{} image: scene under test in the top row; details of the image in the second row; corresponding ratio image in the third row.}
	\label{fig:results_tehran}
	
\end{figure*}

\begin{figure*}[ht]
	\centering
	\begin{tabular}{ccccccc}
		\labNoise & \labProp & \labSardrn &\labIdcnn & \labNoland & \labSarbm & \labFans\\
		\image{accra2_noisy} 	& \image{accra2_prop} 			& \image{accra2_sardrn} 			& \image{accra2_idcnn} 			& \image{accra2_noland} 			& \image{accra2_sarbm3d}			& \image{accra2_fans}\\
    \image{accra2_noisy_d1} & \image{accra2_prop_d1} 		& \image{accra2_sardrn_d1} 			& \image{accra2_idcnn_d1} 		& \image{accra2_noland_d1}			& \image{accra2_sarbm3d_d1}		& \image{accra2_fans_d1}\\
    	& \image{accra2_prop_ratio_d1} 	& \image{accra2_sardrn_ratio_d1} 	& \image{accra2_idcnn_ratio_d1} & \image{accra2_noland_ratio_d1}	& \image{accra2_sarbm3d_ratio_d1}	& \image{accra2_fans_ratio_d1}\\
	\end{tabular}
	\caption{Results on \sent{} image: scene under test in the top row; details of the image in the second row; corresponding ratio image in the third row.}
	\label{fig:results_accra}
	
\end{figure*}

\begin{figure}
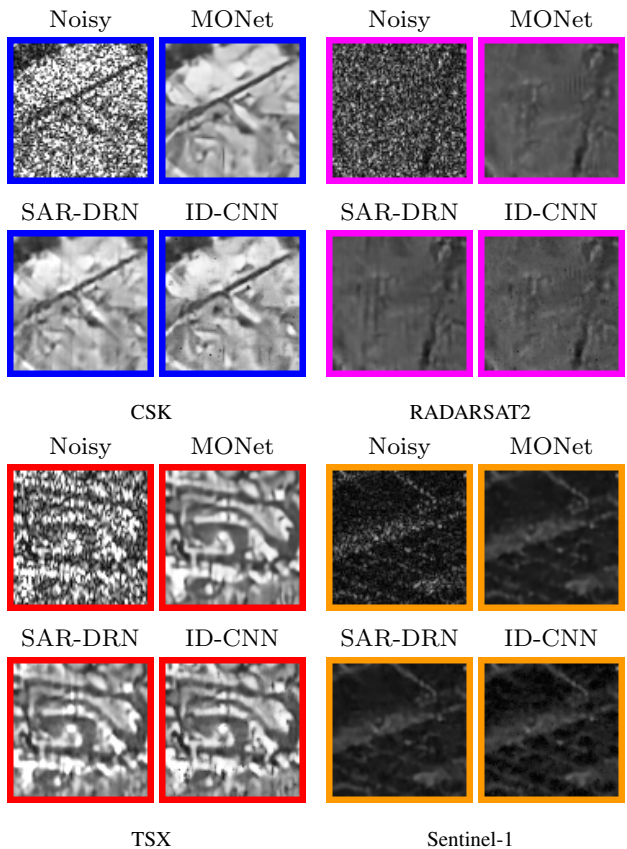


    \centering
    \begin{tabular}{cc}
    \image{pavia_subimg}     & \image{phoenix_subimg} \\
    \labsty{\csk{}}  & \labsty{\radarsat{}}\\
    \image{tehran_subimg}     & \image{accra2_subimg} \\
    \labsty{\tsx{}} & \labsty{\sent{}}\\
    \end{tabular}
    
    \caption{Zoomed detail of blue, magenta, red and orange boxes of Figs. \ref{fig:results_pavia}-\ref{fig:results_accra} from  \csk{}, \radarsat{}, \tsx{} and \sent{} datasets, respectively.}
    \label{fig: zoom}
\end{figure}
In this section we show the results of the proposed algorithm on real SAR data.  Four different test cases have been considered using images acquired by different sensors (COSMO-SkyMed, TerraSAR-X, RADARSAT2, \sent{}), working bands (X and C), acquisition modalities (stripmap and interferometric wide swath), resolutions (three and five meters) and polarization (HH and VH). These has been done in order to show the independence of the achievable results with respect to the considered datasets.

In Figs. \ref{fig:results_pavia}-\ref{fig:results_accra}, the noisy images (first column) and the results on the considered algorithms (columns 2-7) are shown.
In the first row, the results on the whole image are shown. In the second row the results on a particular patch of the whole image is presented. The corresponding ratio images are in the third row. 
Because of the lack of a reference, it is difficult to find a metric that can evaluate fairly the filters given they rely on certain mathematical assumption for the speckle that  is not sure are confirmed in the real SAR image under test. 
For this reason the evaluation of filtering performance mostly relies on visual inspection considering the ability of suppressing noise while preserving objects in the scene.
To this aim, also the ratio images produced by each method are shown.
As noticed in the simulated results, FANS has a good edges preservation but produces over-smoothed results on homogeneous areas. 
NOLAND better preserves spatial details than FANS, but it is still smooth. SAR-BM3D has the best edges and objects preservation among the NL filters, 
but the noise is still present on the filtered images.
Generally, the deep learning solutions try to more suppress the noise compared with the NL approaches.
MONet shows a good trade-off between noise suppression and edges preservation: in homogeneous areas noise is removed without losing many spatial details. 
Moreover, at the same time the edges are quite well preserved. Similar considerations can be done for SAR-DRN and IDCNN but both of them produces some artefacts: the former introduces a vertical texture in all the images and produces disturbed edges, generating less clean images; the latter has a good edges preservation but less suppresses the noise with respect the other two and produce some black spots.

These considerations can be appreciated on the details and on the relative ratio images shown in Figs. \ref{fig:results_pavia}-\ref{fig:results_accra}.

For example, in the COSMO-SkyMed (\csk{}) image of Fig. \ref{fig:results_pavia}, the boundaries of the road are retained quite well from MONet and homogeneous areas are reach of spatial details 
not deleted by the noise removal.

These spatial details barely appear in the NL approaches (except for SAR-BM3D), while edges are well defined.

Moving to the ratio image, it must be recalled that an ideal filter should produce an uncorrelated ratio image: more correlation, more structures are visible in the ratio, worse is the filtering effect.
From the ratio images, the road is more visible for SAR-DRN respect the others: meaning it is heavily filtered and not well preserved. 
The ratio images of IDCNN and MONet are very similar each other with some emergent structure for the former. 
The NOLAND ratio image looks almost uncorrelated but it is characterized by a large granularity typical of generalized smoothness.
Contrary, SAR-BM3D has a ratio image with a very tiny granularity typical of good object preservation but also of a not perfect noise
suppression. 
From the FANS ratio image, it is easily observable a different behaviour of the filter in different areas: 
large granularity on homogeneous areas proving its over smoothing effect, and very tiny granularity in correspondence of 
not homogeneous areas.

Same behavior can be appreciated on the \radarsat{} image in Fig.\ref{fig:results_phoenix}. Even if some structures are more highlighted in the ratio image for MONet, it is still going to have a 
better edges preservation than other methods, except for SAR-BM3D and FANS. 
At the same time, these two solutions still present their limitation: presence of residual noise for SAR-BM3D and over-smoothing for FANS.
In Fig. \ref{fig:results_tehran} the results for the TerrraSAR-X (TSX) image are shown. This image is very challenging for all the solutions, but generally the previous considerations are still valid.
SAR-BM3D is still noisy, FANS tends to over-smooth and NOLAND has a good detail preservation.
The results of MONet and SAR-DRN are similar while ID-CNN produces many artifacts.
Observing the ratio images, proposed solution produces less structures compared with other DL methods, meaning a better preservation of details. It can be noted, in correspondence of urban structures, the appearance of textures within the proposed solutions. Such effect can be expected due invalid fully developed hypothesis for such areas (i.e. extremely heterogeneous areas). This point will be better discussed and analysed in Section \ref{sec:detection}.

The results on \sent{} image are shown in Fig.\ref{fig:results_accra}. The NL solutions have a good edge preservation and generally good noise suppression. SAR-DRN and MONet have similar performance, while ID-CNN suffers on the points characterized by low amplitude value. 

In order to better spot the aforementioned differences among the DL methods, the zooms of the details highlighted in the blue, magenta, red and yellow square boxes of Figs. \ref{fig:results_pavia}-\ref{fig:results_accra} for the correspondent \csk{}, \radarsat{}, \tsx{}, \sent{} datasets  are shown in Fig. \ref{fig: zoom}.

For \csk{} and \radarsat{}{} details, MONet has better edges preservation, SAR-DRN shows smoothness and a vertical texture, ID-CNN is a bit more noisy and presents some black artefacts.

Regarding the zoom of \tsx{},  it can be noted how the proposed MONet tends to enforce smoothness on homogeneous areas but at the same time preserves edges with a slight better quality compared to SAR-DRN. Indeed, the path inside the zoom are better retained respect to IDCNN and SAR-DRN.

The performance of the filters are quite similar on the detail of \sent{} , where ID-CNN shows always some black spots.

In the end, the proposed MONet shows an edge preservation comparable with NL approaches but with better noise suppression resulting in a very good objects and details preservation.
Compared with SAR-DRN and IDCNN, it seems that the depth of the network combined with the use of the defined cost function helps in suppressing the noise and at the same time keeping intact some details such as edges and small object.

In addition to visual comparison, the numerical assessment for each site under test has been carried out.
The M-index is considered, from which the Haralick homogeneity $\homo$, the residual ENL $\resENL$ and the mean of the ratio $\meanratio$ has been extracted. Moreover, the $\kl$ between the pdf of the predicted speckle and the Rayleigh distribution has been reported.

\begin{table}[ht]
	\centering
	\caption{Numerical assessment \csk{}}
	\setlength{\tabcolsep}{5pt}
	\renewcommand\arraystretch{1.3}
	
	\begin{tabular}{|c|ccc|ccc|}
		\hline
		&	\labM & \labH & \labENL & \labMeanRatio& \labKL &\labENLL\\
		\hline
		\labNoise & - & - & - & - & -  &  0.995\\
		\hline                                                       
		\labFans	&16.96			&0.2003			& \first{138} 	& 0.8750  	&0.0224	& \first{36}	\\	
		\labSarbm	&19.81			&0.0275			& 368 			& 0.8790 	&0.1181	& 7	\\	
		\labNoland	&\first{10.61}	&0.0507			& \second{161}	& \second{0.8841} 	&\first{0.0068} & \second{15}\\	
		\hline                                                            
		
		\labIdcnn	&12.64			&0.0021			& 250 			& 0.8758 	&0.0235	& 12	\\	
		\labSardrn	&11.14			&\first{0.0001}	& 222 			& 0.8786 	&0.0393 & \second{15}		\\	
		\labProp	&\second{10.69}	&\second{0.0002}& 213 			& \first{0.8903} 	&\second{0.0178} & 14\\
		
		\hline
		
	\end{tabular}
	\label{tab: numerical pavia}
\end{table}

\begin{table}[ht]
	\centering
	\caption{Numerical assessment \radarsat{}}
	\setlength{\tabcolsep}{5pt}
	\renewcommand\arraystretch{1.3}
	
	\begin{tabular}{|c|ccc|ccc|}

		\hline
		&	\labM & \labH & \labENL & \labMeanRatio & \labKL &\labENLL\\
		\hline
		\labNoise & - & - & - & - & -  &  1\\
		\hline
		\labFans	&	12.40		&	0.1412					&\first{106} &	0.8786	&	\second{0.0206}	&\first{69}\\
		\labSarbm	&	16.42		&	0.0500					&278		 &	\second{0.8788}	&	0.0906	&15		\\
		\labNoland  &	\first{8.75}&	0.0383					&\second{136}&	\first{0.8844}	&	\first{0.0058}	&21\\
		\hline                                                               
		
		\labIdcnn	&	9.52		&	\first{0.0001}			&257		 &	0.8739	&	0.0346			&18\\
		\labSardrn	&	12.90		&	\second{0.0010}			&190		 &	0.8779	&	0.0404			&22\\
		\labProp	&	\second{9.42}		&	\first{0.0001}	&188		 &	0.8677	&	0.0254	&\second{29}\\
		\hline
		
	\end{tabular}
	\label{tab: numerical phoenix}
\end{table}

\begin{table}[ht]
	\centering
	\caption{Numerical assessment TSX - Tehran}
	\setlength{\tabcolsep}{5pt}
	\renewcommand\arraystretch{1.3}
	
	\begin{tabular}{|c|ccc|ccc|}
		\hline
		&	\labM & \labH & \labENL & \labMeanRatio & \labKL &\labENLL\\
		\hline
		\labNoise & - & - & - & - & -  &  0.999\\
		\hline
        \labFans	&	\second{21,96}	&	0.0140			& \second{42}&	0.8583		&	0.2376	            &\first{7}	\\
		\labSarbm	&	56.37			&	0.0882			& 103		 &	\first{0.9026}		&	0.3975		&2	\\
		\labNoland	&	\first{14.07}	&	0.0460			& \first{23} &	0.8944		&	\first{0.0193}	    &\second{6}\\
		\hline                                                                          
		\labIdcnn	&	26.79			&	0.0289			& 50		 &	0.8855		&	0.0767			   &5\\
		\labSardrn	&	23.96			&	\second{0.0006}	& 47		 &	0.8829		&	0.1348			    &5\\
		\labProp	&	24.34			&	\first{0.0001}	& 48		 &	\second{0.9006}	&	\second{0.0651}	&5\\
		\hline
		
	\end{tabular}
	\label{tab: numerical Tehran}
\end{table}
\begin{table}[ht]
	\centering
	\caption{Numerical assessment \sent{}}
	\setlength{\tabcolsep}{4pt}
	\renewcommand\arraystretch{1.3}
	
	\begin{tabular}{|c|ccc|ccc|}
		\hline
		&	\labM & \labH & \labENL & \labMeanRatio& \labKL &\labENLL\\
		\hline 
		\labNoise & - & - & - & - & -  & 1\\
		\hline
		\labFans	 	&16.66 			&0.1465 			&\second{19} &0.8729 			&0.0364 			&\first{30 }\\	
		\labSarbm 	&25.15 			&0.0673 			&44 			&\second{0.8783} &0.1702 			&7 \\	
		\labNoland 	&\first{9.94 }	&0.0170 			&\first{18 }	&0.8776 			&\first{0.0077} 	&\second{16 }\\	
				  \hline     		                      		           	          		                  		    				  		  			  					 		
		\labIdcnn 	&18.39 			&0.0033 			&36 			&\first{0.9100 }	&\second{0.0281 }&12 \\	
		\labSardrn 	&14.02 			&\second{0.0001 }&28 			&0.8731 			&0.0578 			&14 \\	
		\labProp	 	&\second{13.99 }	&\first{0.0000 }	&2 			&0.8741 			&0.0344 			&14 \\
		
		\hline
		
	\end{tabular}
	\label{tab: numerical accra}
\end{table}

Regarding the M-index, NOLAND has always the best value followed by MONet , except for the \tsx{} image, where the second best is FANS.
In order to interpret these results, the three factors $\homo$, $\resENL$ and $\meanratio$, whose M-index is a combination, have been extracted.
Lower is $\homo$, less are the remaining structure and higher is the detail preservation during the noise suppression.
Lower is $\resENL$, the ENLs computed on the ratio image are closer to the ENLs computed to the noisy, meaning a better statistical preservation of the noise.
MONet  shows always the best or the second best value for $\homo$ confirming a better details preservation w.r.t other methods.
Indeed, the other methods produce more artefacts and the ratio images highlight more structures.
Reverse is the situation for the performance on the $\resENL$: MONet  is always surpassed by NOLAND and FANS.

Generally, from the Tabs \ref{tab: numerical pavia}-\ref{tab: numerical Tehran} we can see that DL methods outperform NL methods on $\homo$, but the situation is reverted on $\resENL$. This can be explained by the fact that DL methods are trained under the fully developed hypothesis that is not correct everywhere inside the images, and so the statistical $\resENL$ highlights this characteristic.

Moreover, together with the $\resENL$ the mean value of the ratio images $\meanratio$ have been extracted. The ideal the filter should produce a mean ratio equal to one.
Except for the \radarsat{} where proposed solution reaches the lowest performance, in the \csk{} and \tsx{} it reaches the best and second best performance, respectively, confirming a good quality filtering process.

 For considering the ability in noise suppression, the ENL on homogeneous areas for each image under test has been computed. The selected ares are highlighted in the green boxes, and corresponding ENL of the noisy images are shown in the Tabs \ref{tab: numerical pavia}-\ref{tab: numerical accra}.
Generally, the ENL performance are very close for all the methods, except for FANS that strongly outperforms the others.
This can be explained by the oversmooth behaviour of FANS with respect to the other solutions. 
Regarding the $\kl$, we can see that NOLAND has always the best performance. This thanks to the fact that $\kl$ is included in the
similarity research process.
Proposed MONet  has the second best performance on \csk{} and on \tsx{}, while on \radarsat{} it reaches the third one.
Naturally the $\kl$ results are affected by presence of not homogeneous areas and so they are rather general.
It is worth to notice that among the DL methods, MONet  has always the best $\kl$ index. 
This means that using a statistical term as $\lkl$ gives the network an added useful statistical information that can not be acquired only by the data.

\subsection{ Identification of Not Fully Developed Areas: Validation}
It is worth to notice that our network is trained under the fully developed hypothesis and the use of $\ledge$ aims in preserving objects, details and strong scatterers where that hypothesis is not valid anymore. These points strongly appear in the ratio images produced by the proposed method.
 As described in section \ref{sec:detection}, an identification step allows to isolate such points leaving the user the possibility to decide the filtering policy. 

\begin{figure}[ht]
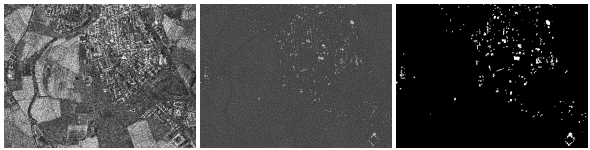

	\centering
	\begin{tabular}{ccc}
		\image{pavia_noisy_det} & \image{pavia_prop_ratio_det} & \image{pavia_prop_mask}\\
	\end{tabular}
	\caption{Result of the detection process on \csk{} dataset, from left to right: SAR image; ratio image produced by proposed CNN; detection result}
	\label{fig:pavia-detec}
	
\end{figure}
\begin{figure}[ht]
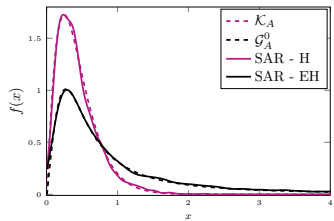

	\centering
	\image{pavia_distr}
	\caption{Comparison between the distributions of two different areas of SAR image  (solid) with the theoretical ones (dashed): black solid curve represents the distribution of extremely heterogeneous detected points on SAR image (SAR-EH); the black dashed curve is the theoretical  $\mathcal{G}_A^0(-0.5,0.145,1)$; magenta solid curve represents the distribution of an heterogeneous area of SAR image (SAR-H); magenta dashed is the theoretical $\mathcal{K}_A(2,7.5,1)$}
	\label{fig: pavia-distribution}
	
\end{figure}

 
The validation of this procedure is performed in the following only on the \csk{} dataset, however similar results can be achieved using the other datasets. 
In Fig.\ref{fig:pavia-detec} the detection map of not fully developed points are shown for \csk{}. 
In Fig. \ref{fig: pavia-distribution}, it is shown how the detected points on the SAR image (SAR Extremely Heterogeneous points, SAR-EH) generate a pdf (solid magenta curve) that well fits the theoretical distribution of  $\mathcal{G}_A^0(\alpha,\gamma,L)$ (dashed magenta)   of \cite{Frery1997} as the distribution that better describes such areas. The parameters are empirically estimated as $(\alpha=-0.5,\gamma=0.145,\lambda=0,L=1).$

At the same time, we estimated the distribution of the SAR image in the remaining points (SAR Heterogeneous points, SAR-H): this fits the $\mathcal{K}_A(\alpha,\lambda,L)$ distribution, meaning that all the remaining part of the image belongs to heterogeneous areas. The parameters are empirically set as $(\alpha=2,\gamma=0,\lambda=7.5,L=1).$

\begin{figure}
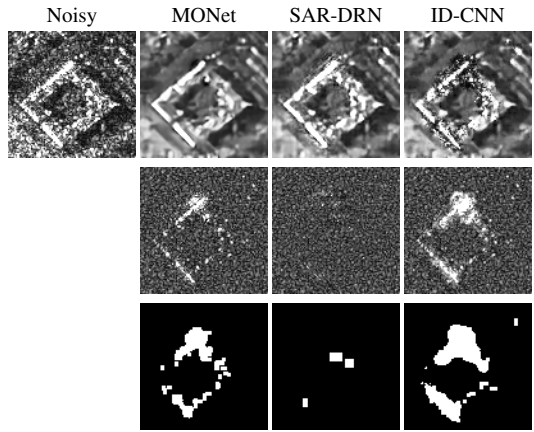

	\centering
	\begin{tabular}{cccc}
		\labNoise	& \labProp & \labSardrn & \labIdcnn\\
		\image{pavia_noisy_det_d1}  & \image{pavia_prop_det_d1} & \image{pavia_sardrn_det_d1} & \image{pavia_idcnn_det_d1}\\
							& \image{pavia_prop_ratio_det_d1} & \image{pavia_sardrn_ratio_det_d1} & \image{pavia_idcnn_ratio_det_d1}\\
							& \image{pavia_prop_mask_d1} & \image{pavia_sardrn_mask_d1} & \image{pavia_idcnn_mask_d1}\\
								   
	\end{tabular}
	\caption{Comparison of the detection of strong scatterers between the proposed MONet , SAR-DRN and ID-CNN. In the top noisy and ratio images are shown. In the bottom the relative detection.}
	\label{fig: detection_comparison}
\end{figure}

This confirms the fact that our CNN is able to detect the points belonging to the extremely heterogeneous areas directly from the ratio image. 
Naturally, this issue is in common with all the CNN that use training data simulated under the fully developed hypothesis.
So this procedure could be extended also to the other methods like ID-CNN and SAR-DRN.
In Fig. \ref{fig: detection_comparison} a patch of \csk{} is shown with relative detection for the DL methods.
First of all, it is important to note the different behavior of three CNNs on strong scatterers: MONet  try to isolate the objects by preserving the edges and at the same time produce a strong structure in the ratio; contrary SAR-DRN try to less filter these elements but some distortion are visible both in the filtered image and in the ratio. ID-CNN produces many artefacts not only in correspondence of the scatterers but also in its neighborhood.

By visually inspecting the amplitude image and the detection results, it is evident that many point scatterers are not correctly identified by SAR-DRN, many false identification are present  in ID-CNN while the MONet  seems the most reliable.

%% file: 4.conclusion.tex
\section{Conclusion}
\label{sec:conclusion}
In this paper a CNN for SAR despeckling trained on simulated data has been proposed. The non linearity introduced by the seventeen layers are crucial for features extraction while skip connections are used for avoiding the vanishing gradient problem.
Beyond the proposed architecture, the main focus is dedicated to the definition of a multi-objective cost function composed of three terms: $\lmse$ , $\lkl$, $\ledge$. The combination of these three terms allows the preservation of spatial details, statistical properties, edges and identification of strong scatterers.
An ablation study proves how the combination is crucial for taking care of these aspects simultaneously.
Experimental validation, both on simulated and real data, show the advantages on including these SAR image properties in the cost function.

The performance on simulated images show an improvement with respect to the state of art, mainly on edges and details preservation. This is confirmed also in real SAR images where the results present good noise rejection, edges preservation and absence of artefacts.
This means a more clear filtered images with well retained edges and objects. 

Moving to real data, MONet based on the fully developed hypothesis and cascade of convolutions cannot directly handle the point scatterers, but differently from other CNN based solutions (such as SAR-DRN and ID-CNN)  it is able to identify them. Hence, the knowledge of those pixels allows the final user to decide how to process them (left unfiltered, define a specific statistical based filter,  using a multitemporal approach, etc…).

Being a DL based method, once the network training is performed the computational time is limited. Further works will address the possibility of adapting the filter to multilook, multitemporal and multichannel SAR images.

%% file: 5.appendix.tex
\section{On the impact of the cost function}
\label{sec:appendix}
In this section an ablation study has been carried out in order to assess the impact of the defined cost function.
The cost function is given by a combination of the terms in Eq. \ref{eq: cost}.
In order to compare the performance and the impact of these three terms, the same architecture is trained on same dataset with a cost function composed 
once only of the $L_2=\lmse$ term , once of the combination $L_{kl}=\lmse + \lambda_{kl} \lkl$, and once with the combination $L_\nabla=\lmse + \lambda_{\nabla} \ledge$.
These solutions are compared with the proposed method.
\begin{table}[h]
	\centering
	\caption{Numerical Assessment on Simulated Dataset for different cost functions: the value are averaged on the whole simulated testing dataset composed of 100 images. From top to bottom: network trained with $L_2$, $L_{kl}$, $L_\nabla$, $\lcost $}
	\setlength{\tabcolsep}{5pt}
	\begin{tabular} {l|ccc}
		
		& 	\labSSIM & \labSNR & \labMSE \\
		\hline
		\hline
		$L_2$		&	0,7509 &	8,8514 &	0,0040	\\
		$L_{kl}$		&	0,7514 &	8,8543 &	0,0039	\\
		$L_{\nabla}$	&	0,7512 &	8,8585 &	0,0039  \\
		$\lcost$	&	0,7510 &	8,8555 &	0,0039	\\   
		\hline
		\end{tabular}
	
	\label{tab: ablation}
\end{table}
In Tab. \ref{tab: ablation}, we summarize the numerical assessment on the same testing dataset of Sec. \ref{sec: simulated results}.
The results are almost the same for each solution, like there is no difference in introducing such terms in the cost function.
It seems that  $\lmse$ is enough for the despeckling. However,  these are average metrics that do not take into account the details that make the difference between one solution and an other. Moving to real data, things largely change. 

\begin{figure}[h]
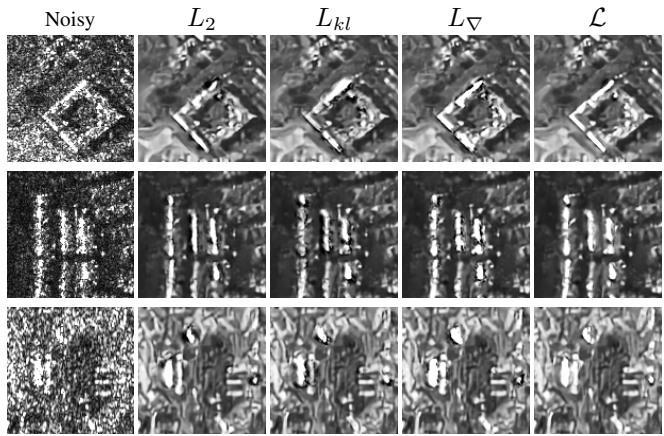

	\centering
	\begin{tabular}{ccccc}
		\labNoise & $L_2$ & $L_{kl}$ & $L_\nabla$ & $\lcost$ \\
		\image{pavia_noisy_abl} & \image{pavia_l2_abl} & \image{pavia_l2kl_abl}&\image{pavia_l2edge_abl}&\image{pavia_prop_abl}\\
		\image{phoenix_noisy_abl} & \image{phoenix_l2_abl} & \image{phoenix_l2kl_abl}&\image{phoenix_l2edge_abl}&\image{phoenix_prop_abl}\\
		\image{tehran_noisy_abl} & \image{tehran_l2_abl} & \image{tehran_l2kl_abl}&\image{tehran_l2edge_abl}&\image{tehran_prop_abl}\\
		
	\end{tabular}
	\caption{Details for the different cost function}
	\label{fig: results ablation}
\end{figure}

In Fig. \ref{fig: results ablation}, a detail for each dataset \csk, \radarsat and \tsx are shown. It can be noted how important is the impact of the cost function.
Starting from $L_2$ that try to preserve spatial details, the use of the KL divergence in $L_{kl}$ helps in filtering the homogeneous areas but we lose information on strong scatterers producing smoothing effect. In addition, the $L_\nabla$ try to preserve edges but does not consider the speckle properties and tends to create strange artefacts in the neighbourhood of the strong scatterers. The proposed cost function $\lcost$ is able to balance all these effects and to give best compromise.
The use the $\lcost$ allows the filter the image balancing at the same time the statistical properties of the noise and the details preservation.

\section*{Acknowledgments}
The CSK data have been provided by ASI within the project COSMO-SkyMed SAR data - Contract n. I/065/09/0. The TSX data have been have been provided by DLR within the framework of the Project MTH3649. The RS2 data are free available at https://mdacorporation.com. RADARSAT-2 Data and Products © MacDONALD, DETTWILER AND ASSOCIATES LTD. – All Rights Reserved.  RADARSAT is an official mark of the Canadian Space Agency”. Sentinel data are provided by ESA - Copernicus Sentinel data 2018, processed by ESA.